\documentclass[]{jpsj3}
\usepackage{txfonts}

\newcommand{\Tss}[3]{ \ensuremath{#1 _{\mathrm{#2}} ^{\mathrm{#3}} }}
\newcommand{\Tsub}[2]{\ensuremath{#1_{\mathrm{#2}}}}

\newcommand{\Tk}[1]{\Tsub{\kappa}{#1}}

\newcommand{\Tkax}[1]{\ensuremath{\kappa _{#1}}}
\newcommand{\Tl}[1]{\Tsub{l}{#1}}

\newcommand{\THax}[1]{\ensuremath{H _{#1}}}

\newcommand{\TT}[1]{\Tsub{T}{#1}}

\newcommand{\TJ}[2]{\Tss{J}{#1}{#2}}
\renewcommand{\surname}[1]{\TextRed{#1}}
\renewcommand{\THax}[1]{\ensuremath{H _{\parallel #1}}}
\newcommand{\TextRed}[1]{{#1}}
\newcommand{\TextBlue}[1]{{#1}}
\newcommand{\TextOliveGreen}[1]{{#1}}
\newcommand{\TRevA}[1]{{#1}}
\newcommand{\TRevB}[1]{{#1}}
\newcommand{\TRevC}[1]{{#1}}
\newcommand{\TRevD}[1]{{#1}}
\newcommand{\TRevE}[1]{{#1}}

\title{Magnetic State of \TextBlue{the} Geometrically Frustrated Quasi-One-Dimensional Spin System Cu$_3$Mo$_2$O$_9$ Studied by Thermal Conductivity}

\author{\name{Koki \surname{Naruse}}$^1$\thanks{Present address: Institute for Materials Research (IMR), Tohoku University, Sendai 980-8577, Japan}, \name{Takayuki \surname{Kawamata}}$^1$\thanks{E-mail: tkawamata@teion.apph.tohoku.ac.jp}, \name{Masumi \surname{Ohno}}$^1$, \name{Yoshiharu \surname{Matsuoka}}$^1$,\\ \name{Masashi \surname{Hase}}$^2$,
\name{Haruhiko \surname{Kuroe}}$^3$, \name{Tomoyuki \surname{Sekine}}$^3$, \name{Kunihiko \surname{Oka}}$^4$, \name{Toshimitsu \surname{Ito}}$^4$, \name{Hiroshi \surname{Eisaki}}$^4$, \name{Takahiko \surname{Sasaki}}$^5$, and \name{Yoji \surname{Koike}}$^1$}
\inst{$^1$Department of Applied Physics, Tohoku University, Sendai 980-8579, Japan \\
$^2$National Institute for Materials Science (NIMS), Tsukuba 305-0047, Japan \\
$^3$Department of Physics, Sophia University, Tokyo 102-8554, Japan \\
$^4$National Institute of Advanced Industrial Science and Technology (AIST), Tsukuba 305-8501, Japan\\
$^5$Institute for Materials Research (IMR), Tohoku University, Sendai 980-8577, Japan}

\abst{
We have measured the thermal conductivity of \TextBlue{the geometrically frustrated quasi-one-dimensional spin system Cu$_3$Mo$_2$O$_9$} in magnetic fields. 
\TextBlue{A} contribution of the thermal conductivity due to spin\TextBlue{s} has been observed in the thermal conductivity along the spin chains. 
The thermal conductivity due to phonons, $\kappa_{\rm phonon}$, \TextBlue{has been found to decrease by the application of} a magnetic field, \TextBlue{which has been explained as being due to the reduction in the} spin gap \TextBlue{originating} from the spin-\TextBlue{singlet} dimers. 
\TextBlue{Moreover, }
it has been found that $\kappa_{\rm phonon}$ \TextBlue{increases with increasing field in high fields above $\sim 7$~T at low temperatures.} 
\TRevB{This suggests} the existence of a novel \TextBlue{field-induced} spin state \TRevB{and} is discussed in terms of the \TextBlue{possible spin-}chirality ordering in \TextBlue{a} frustrated Mott insulator. 
}

\begin{document}
\maketitle

\newpage

\section{Introduction}

\TextBlue{The} thermal conductivity in \TRevD{low-dimensional} quantum spin systems has attracted \TextBlue{great} interest\TextBlue{,} because \TextBlue{a large amount of} \TextRed{thermal conductivity due to \TextBlue{spins, namely,} magnetic excitations, $\kappa_{\rm spin}$, has been observed} along the direction where \TextBlue{the} antiferromagnetic (AF) exchange interaction is strong.  
\TRevD{In the AF spin-chain systems Sr$_2$CuO$_3$\cite{Takahashi_2006, Kawamata_2008, Sologubenko_2001} and SrCuO$_2$\cite{Sologubenko_2001, Kawamata_2010} and the two-leg spin-ladder system Sr$_{14}$Cu$_{24}$O$_{41}$ \cite{Kudo:JLTP117:1999:1689,Kudo:JPSJ70:2001:437,Sologubenko:PRL84:2000:2714,Hess:PRB64:2001:184305,Naruse:SSC154:2013:60}, for example, the thermal conductivity due to spinons and magnons, which are magnetic excitations in these systems, has been observed, respectively.} 
In addition, 
the thermal conductivity \TextBlue{has attracted considerable interest, because it is closely related to the magnetic state.} 
\TextBlue{That is, the thermal conductivity exhibits a marked change according to the change in the magnetic state, owing to the marked change in the scattering of heat carriers by magnetic excitations}. 
\TextRed{In the spin-Peierls system CuGeO$_3$ \cite{Ando_1998,Salce1998:127} and the two-dimensional spin-dimer system SrCu$_2$(BO$_3$)$_2$ \cite{Kudo_2001, Hofmann_2001}, 
the \TextRed{thermal conductivity due to phonons}, $\kappa_{\rm phonon}$, has been \TextBlue{found to be enhanced at low temperatures} below \TextBlue{the} temperature \TextBlue{comparable} to \TextBlue{the} spin-gap \TextBlue{energy owing to the reduction in the phonon-spin scattering rate} and \TextBlue{to be} suppressed by the application of a magnetic field} because of the \TextBlue{reduction} in the spin gap \cite{Kudo_2001, Hofmann_2001}. 
Furthermore, 
\TextRed{a marked enhancement of the thermal conductivity has been observed at low temperatures below \TextBlue{the} AF transition temperature, \TT{N}, in several AF spin systems \cite{Slack1958,Slack1961,Aring1967,Lewis1973,Kudo_2003,Sologubenko2003a,Zhao2012}.} 
\TextRed{In \TextBlue{the} frustrated spin system Dy$_2$Ti$_2$O$_7$, \TextBlue{recently, the thermal conductivity has been found to be} affected by \TextBlue{the} change in \TextBlue{the} state of magnetic monopoles, which are magnetic excitations in this system.\cite{Kolland_2013,Fan_2013}} 
\TextBlue{Accordingly}, the thermal conductivity is \TextBlue{recognized as a} very useful \TextBlue{probe to detect a} change in \TextBlue{the} magnetic state and \TextBlue{a} phase transition.

\begin{figure}[b]
\begin{center}
\centering
\includegraphics[width=0.7\linewidth]{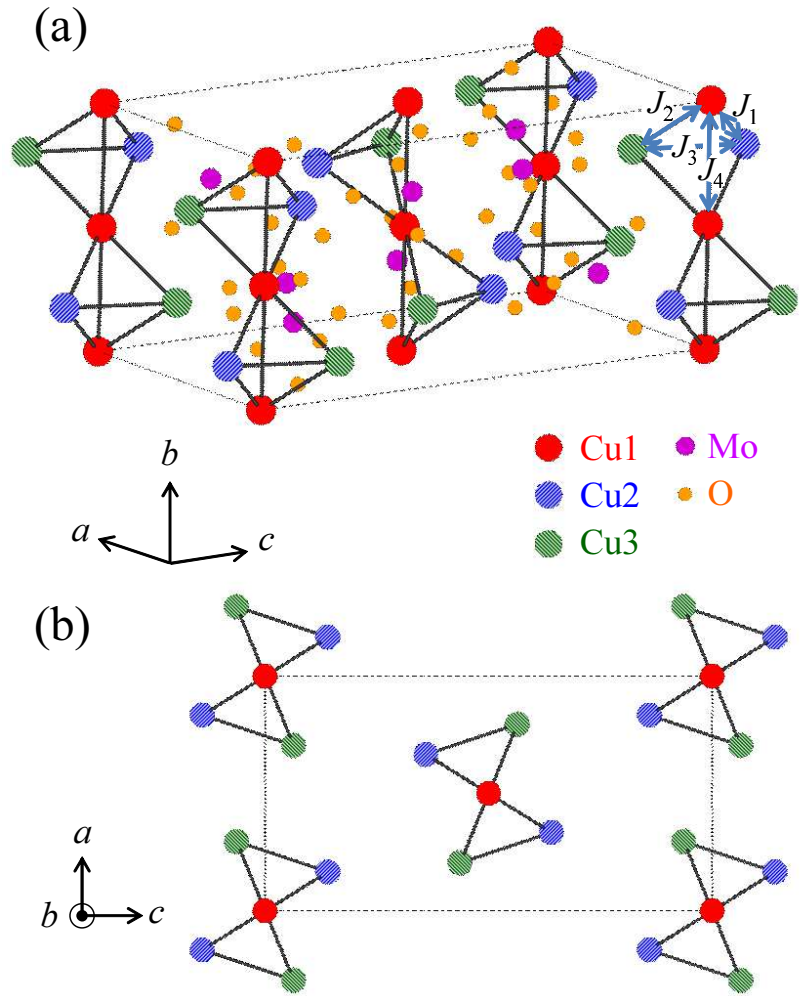}
\vspace{5mm}
\caption{(Color online) 
(a) \TextBlue{Crystal structure of Cu$_3$Mo$_2$O$_9$.} 
\TextBlue{Distorted} tetrahedral spin-chain\TextBlue{s} run along the $b$-axis. 
(b) \TextBlue{Crystal structure} viewed from the $b$-axis. 
\TextBlue{Dashed lines} indicate the unit cell \TextBlue{containing two distorted tetrahedral spin-chains}. }
\label{fig01}
\end{center}
\end{figure}


\TextBlue{The compound Cu$_3$Mo$_2$O$_9$ is a \TRevD{quasi-one-dimensional} spin system with the quantum spin number $S = 1/2$ of Cu$^{2+}$ ions.} 
\TextBlue{As shown in Fig. 1(a),} 
\TRevA{distorted tetrahedral spin-chains composed of spin chains of Cu1 and spin dimers of Cu2 and Cu3 run along the $b$-axis.}  
\TextRed{The spin chains are arranged in the $ac$-plane as shown in Fig. \ref{fig01}(b). }
\TextRed{Cu$^{2+}$} \TextBlue{spins} interact \TextBlue{with} one another by AF \TextRed{superexchange} interactions, 
\TextRed{whose} 
magnitude has been estimated 
from the inelastic \TextBlue{neutron-}scattering \TextBlue{experiment} as follows \cite{Kuroe_2010,Kuroe_2011_PRB}. 
\TRevB{
Both the interaction between Cu1 and Cu2, \TJ{1}{}, and that between Cu1 and Cu3, \TJ{2}{}, are $\sim 19$ K. 
\TRevC{The intradimer interaction between Cu2 and Cu3, \TJ{3}{}, which is equal to the spin-gap energy, $\Delta$, of the spin dimers, is $\sim 67$~K.} 
The intrachain interaction between Cu1's, \TJ{4}{}, is $\sim 46$~K.} 
\TextBlue{The} 
interchain interaction \TextBlue{is as negligibly weak as} $\sim$ 2.2 K. 

%
%
\TextBlue{The} 
\TextRed{magnetization and specific heat measurements} 
\TextBlue{have revealed that} 
Cu$_3$Mo$_2$O$_9$ undergoes \TextBlue{an} AF \TextBlue{transition} accompanied by weak ferromagnetism (WF) due to \TextBlue{the} Dzyaloshinsky-Moriya interaction at 8 K \cite{Hamasaki_2008}. 
\TextRed{In the AF ordered state}, 
\TRevB{ the dispersion branch \TRevD{of magnetic excitations of} the spin \TRevD{dimers} remains together with that \TRevD{of} the AF order \cite{Kuroe_2010,Kuroe_2011_PRB} and} 
the direction of Cu1 spins is \TextRed{almost} parallel to the $b$-axis 
\TextBlue{but} is slightly canted from the $b$-axis \cite{Hamasaki_2008}. 
\TextBlue{Although} 
\TextRed{canted \TextBlue{components of the} magnetic moments 
\TextBlue{are in disorder in zero field, they are ordered} by the application of }
\TextBlue{a} magnetic field of 0.1 T along the $a$-axis and of 0.8 T along the $c$-axis. 
\TextRed{In the AF ordered state, furthermore,} 
\TextBlue{it has been found from dielectric constant and magnetization measurements that} 
Cu$_3$Mo$_2$O$_9$ shows magnetic 
and ferroelectric \TextBlue{orders simultaneously} without any magnetic superlattice formation,\TextRed{\cite{Kuroe_2011_JPSJ}}
\TextBlue{which has been understood as being due to} 
the \TextBlue{possible} charge redistribution in \TextBlue{a} frustrated Mott insulator \cite{Kuroe_2011_JPSJ, Bulaevskii_2008, Khomskii_2010}. 
The direction of \TextBlue{the} spontaneous electric polarization changes 
\TextRed{from the $c$-axis to the $a$-axis by the application of }
\TextBlue{a} magnetic field of \TextRed{$\sim$ 8~T} along the $c$-axis,\TextRed{\cite{Kuroe_2011_JPSJ}} \TextBlue{which} 
\TextRed{has been also observed in \TextBlue{the} electron-spin-resonance spectrum \TextBlue{of} the powder sample\cite{Okubo_2010}. } 
\TextBlue{At present, the phase diagram of Cu$_3$Mo$_2$O$_9$ in magnetic fields at low temperatures is as shown in Fig. 2.\cite{Hamasaki_2010,Kuroe_2011_JPSJ,Hosaka_2012} } 
Nevertheless, the magnetic state of Cu$_3$Mo$_2$O$_9$ has not \TextBlue{yet} been clarified completely.
Accordingly, we have measured the thermal conductivity 
\TextBlue{of single-crystal Cu$_3$Mo$_2$O$_9$ in magnetic fields,} 
in order to investigate the magnetic state of Cu$_3$Mo$_2$O$_9$ 
\TextBlue{as well as the existence of \Tk{spin}}.


\section{Experimental}
Single crystals of Cu$_3$Mo$_2$O$_9$ were grown by the continuous solid-state crystallization method \cite{Oka_2011}.
Thermal conductivity measurements were carried out by the conventional steady-state method.
One side of a rectangular single-crystal, whose typical dimensions were about 5 $\times$ 1 $\times$ 1 mm$^3$, was anchored on \TextBlue{a} heat sink \TextBlue{of copper} with indium solder.
A chip-resistance of 1~k$\Omega$ (Alpha Electronics MP1K000) was attached as a heater to the opposite side of the single crystal with GE7031 vanish.
The temperature difference across the crystal (0.03--0.4~K) was measured with two Cernox thermometers (Lake Shore Cryotronics CX-1050-SD).
The accuracy of the absolute value of the thermal conductivity was $\pm$10$\%$ mainly due to the uncertainty of the sample geometry.
Magnetic field\TextBlue{s} up to 14~T were applied parallel to the principal crystallographic axes. 

\begin{figure}[b]
\begin{center}
\includegraphics[width=1.0\linewidth]{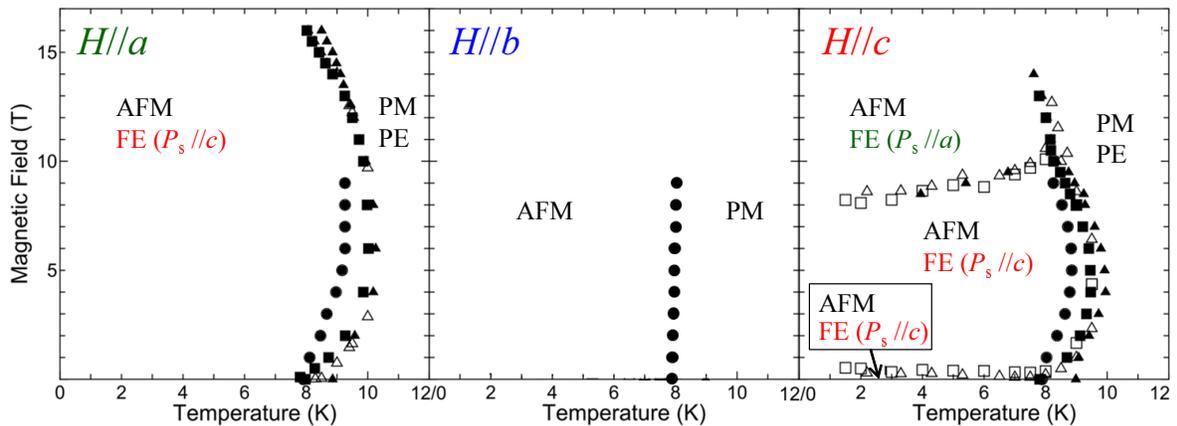}
\caption{(Color online) \TextBlue{Phase diagram of Cu$_3$Mo$_2$O$_9$ in magnetic fields along the principal crystallographic axes at low temperatures.\cite{Hamasaki_2010,Kuroe_2011_JPSJ,Hosaka_2012} }  
\TextOliveGreen{AFM, PM, FE, and PE indicate the antiferromagnetic, paramagnetic, ferroelectric, and paraelectric phases, respectively. 
$P_{\mathrm{s}}$ indicates the spontaneous electric polarization. 
Triangles, squares, and circles were determined from the dielectric constant measurements along the $a$- and $c$-axes and specific heat measurements, respectively. 
Open and solid symbols were obtained from the data of magnetic-field and temperature dependences, respectively. }}
\label{PD}
\end{center}
\end{figure}

\section{Results and Discussion}

\begin{figure}[b]
\begin{center}
\includegraphics[width=0.4\linewidth]{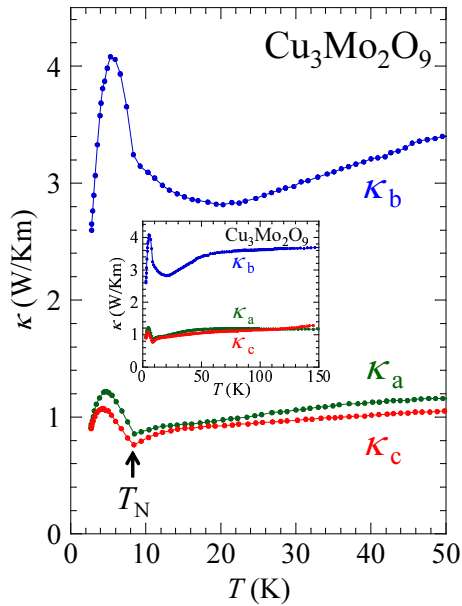}
\vspace{8mm}
\caption{(Color online) Temperature dependence of the thermal conductivity along the $a$-, $b$-, and $c$-axes, $\kappa_{a}$, $\kappa_{b}$, and $\kappa_{c}$, for \TextBlue{Cu$_3$Mo$_2$O$_9$} single crystals \TextRed{in zero field, respectively}. 
\TextBlue{The inset shows the temperature dependences of \Tkax{a}, \Tkax{b} and \Tkax{c} in a wide temperature-range up to 150~K. 
The arrow indicates the antiferromagnetic transition temperature, \TT{N}. }
}
\label{fig02}
\end{center}
\end{figure}

Figure \ref{fig02} shows the temperature \TextRed{dependence} of the thermal conductivity along the $a$-, $b$-, and $c$-axes, $\kappa_{a}$, $\kappa_{b}$, and $\kappa_{c}$, 
\TextRed{of Cu$_3$Mo$_2$O$_9$, respectively}.
\TextRed{It is found that $\kappa_{a}$ and $\kappa_{c}$ perpendicular to the spin chains} 
\TextBlue{are} similar \TextBlue{to each other} 
\TextRed{and monotonically } 
decrease with decreasing temperature \TextRed{\TextBlue{down} to} \TextBlue{$T_{\rm N}~=~8$~K}. 
\TextRed{Although $\kappa_{b}$ parallel to the spin chains also decreases with decreasing temperature \TextBlue{down} to $\sim 20$~K, }
\TextBlue{on the other hand,} 
\TextRed{$\kappa_{b}$ }
increases with decreasing temperature \TextBlue{from $\sim 20$~K down} \TextRed{to $T_{\rm N}$}. 
\TextBlue{Both} $\kappa_{a}$, $\kappa_{b}$ and $\kappa_{c}$ increase suddenly 
\TextBlue{just below \TT{N} with decreasing temperature} 
and exhibit \TextBlue{a} peak at approximately 5 K. 
\TextRed{In nonmagnetic insulators, \TextBlue{$\kappa_{\rm phonon}$ typically increases with decreasing temperature from room temperature} and shows a peak at \TextBlue{a low temperature} around 10 K.}  
\TRevD{In spin-gap systems, moreover, thermal conductivity typically increases with decreasing temperature at low temperatures below the temperature comparable to the spin-gap energy.}  
\TRevD{Taking into account the observation of the dispersion branch of magnetic excitations of the spin dimers,\cite{Kuroe_2010,Kuroe_2011_PRB}}  
\TextRed{therefore, the monotonic decrease with decreasing temperature \TextBlue{at high temperatures} implies }
that
the mean free path of phonons, $l_{\rm phonon}$, is \TextRed{strongly} suppressed \TextBlue{probably} \TextRed{by magnetic fluctuations due to the spin frustration.} 
\TextBlue{The} sudden increases \TextBlue{in} 
\TextRed{$\kappa_{a}$, $\kappa_{b}$, and $\kappa_{c}$ \TextBlue{just} below $T_{\rm N}$} 
are inferred to be due to the increase in $l_{\rm phonon}$ 
\TextRed{owing to the marked reduction in the phonon-spin scattering rate caused by the development of the \TextBlue{AF} long-range order, }
\TextRed{as observed in several antiferromagnets \cite{Slack1958,Slack1961,Aring1967,Lewis1973,Kudo_2003,Sologubenko2003a,Zhao2012}. }

\TextRed{It is found that} 
the magnitude of $\kappa_{b}$ is larger than \TextBlue{those} of $\kappa_{a}$ and $\kappa_{c}$. 
Furthermore, \TextRed{only $\kappa_{b}$ \TextBlue{increases with decreasing temperature at temperatures between $\sim$} 20~K \TextBlue{and} $T_{\rm N}$}, 
\TextBlue{which can hardly be explained as being due to} 
the anisotropy of $\kappa_{\rm phonon}$. 
\TextRed{Therefore, \TextBlue{these} anisotropic \TextBlue{behaviors} of the thermal conductivity \TextBlue{are reasonably attributed to the} contribution of $\kappa_{\rm spin}$ to $\kappa_{b}$,} 
\TRevE{because magnetic excitations can carry heat along the $b$-axis where the magnetic correlation is developed at low temperatures below \TJ{4}{}.} 
\TRevE{Such anisotropic contribution of $\kappa_{\rm spin}$ has been observed in several low-dimensional spin systems.\cite{Takahashi_2006, Kawamata_2008, Sologubenko_2001, Kawamata_2010,Kudo:JLTP117:1999:1689,Kudo:JPSJ70:2001:437,Sologubenko:PRL84:2000:2714,Hess:PRB64:2001:184305,Naruse:SSC154:2013:60,Miike_1975,Nakamura_1991,Matsuoka_2014,Sologubenko2003a,Parfen_2004,Uesaka_2010,Kawamata_2014} }

\begin{figure}[t]
\begin{center}
\includegraphics[width=1.0\linewidth]{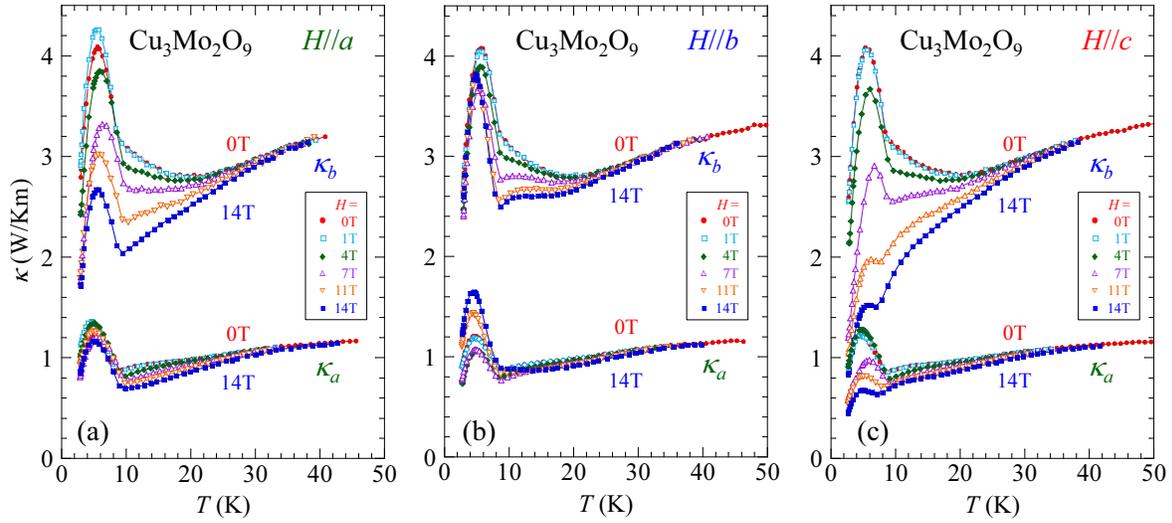}
\vspace{3mm}
\caption{
(Color online) 
Temperature dependence of the thermal conductivity along the $a$- and $b$-axes, $\kappa_a$ and $\kappa_b$, \TextRed{respectively,} for \TextBlue{Cu$_3$Mo$_2$O$_9$} single crystals in  magnetic fields \TextRed{parallel to the (a) $a$-, (b) $b$-, and (c) $c$-axes. }}
\label{fig03}
\end{center}
\end{figure}

\TextRed{Figure \ref{fig03} shows} the temperature \TextRed{dependences} of $\kappa_{a}$ and $\kappa_{b}$ \TextBlue{of Cu$_3$Mo$_2$O$_9$} in magnetic fields along the $a$-, $b$-, and $c$-axes, \TextRed{\THax{a}, \THax{b}, and \THax{c}, respectively}. 
\TextRed{It is found that} \TextBlue{both} 
$\kappa_{a}$ and $\kappa_{b}$ decrease with increasing field 
\TextBlue{at low temperatures} \TextRed{below} \TextBlue{$\sim$}~40~K. 
The decrease \TextBlue{in} $\kappa_{a}$ 
\TextRed{by the application of a magnetic field }
indicates the 
\TextRed{decrease} \TextBlue{in} $l_{\rm phonon}$ 
\TextRed{due to \TextBlue{the} increase in \TextBlue{the phonon-spin scattering rate, namely, the enhancement of} the scattering } \TextBlue{of phonons} 
by magnetic excitations, 
\TextRed{because \TextBlue{the} contribution of $\kappa_{\rm phonon}$ is dominant in $\kappa_{a}$ perpendicular to \TextBlue{the} spin chains} \TextBlue{and the contribution of \Tk{spin} is negligible.} 
\TextRed{It is \TextBlue{known} in spin-gap systems that} 
$\kappa_{\rm phonon}$ is enhanced 
\TextRed{below \TextBlue{the} temperature comparable to \TextBlue{the} spin-gap energy}, 
\TRevD{owing to the marked decrease in the number of magnetic excitations.}  
\TRevD{Moreover, 
the enhancement of $\kappa_{\rm phonon}$ is suppressed by the application of a magnetic field \cite{Ando_1998,Kudo_2001,Hofmann_2001},} 
\TRevD{owing to the increase in the number of magnetic excitations because of the reduction in the spin gap.}  
In the magnetic dispersion of Cu$_3$Mo$_2$O$_9$, there is a \TRevD{flat branch} of magnetic excitations of the spin dimers \cite{Kuroe_2010,Kuroe_2011_PRB}.  
\TRevD{Such a flat magnetic} branch is expected to scatter phonons strongly, because the momentum conservation law is easily satisfied in the phonon-spin scattering process.  
\TRevD{Surely, \Tk{phonon} is suppressed owing to the disorder of the AF correlation induced by the application of the magnetic field in AF spin-chain systems.}  
\TRevD{However, since the magnetic dispersion branch in AF spin-chain systems is dispersive, it is not easy to satisfy both the momentum and energy conservation laws in the phonon-spin scattering process.}  
\TRevD{Therefore, magnetic excitations of the spin dimers are expected to scatter phonons stronger than those of the AF spin chains.} 
\TRevD{Furthermore, considering that }
the temperature below which the suppression by the application of a magnetic field \TRevB{is observed} is comparable to $\Delta = \TRevB{67}$~K \TRevB{\cite{Kuroe_2010,Kuroe_2011_PRB}}, the suppression of not only \Tkax{a} but also \Tkax{b} by the application of a magnetic field is interpreted as being caused by the enhancement of the phonon-spin scattering due to the reduction in the spin gap. 
\TextRed{However, \TextBlue{neither} enhancement of \TextBlue{\Tkax{a}, \Tkax{b}, nor \Tkax{c}} is observed \TextBlue{in zero field below $\sim 40$~K \TRevB{comparable to $\Delta$}.}} 
\TextRed{This may indicate that phonon\TextBlue{s are} strongly scattered by magnetic fluctuations due to \TextBlue{the} spin frustration even \TRevB{at low temperatures} \TRevB{below $\Delta$}.} 
\TextRed{Furthermore,} the decrease \TextBlue{in} $\kappa_{b}$ \TextBlue{by the application of a magnetic field} is \TextBlue{more marked} than that in $\kappa_{a}$. 
This indicates that not only $\kappa_{\rm phonon}$ but also $\kappa_{\rm spin}$ decrease\TextBlue{s} 
\TextRed{by \TextBlue{the} application of a magnetic field, }
\TextRed{because \TextBlue{there exists the contribution of} $\kappa_{\rm spin}$  to $\kappa_{b}$ parallel to \TextBlue{the} spin chains \TextBlue{in zero field as described above}. }
\TextBlue{It is reasonable that \Tk{spin} is affected by magnetic fields up to 14~T, because $\TJ{4}{} \sim 46$~K is not much larger than the energy of a magnetic field of 14~T.} 
\TRevE{Namely, magnetic excitations carrying heat are scattered by the disorder of the antiferromagnetic correlation along the $b$-axis induced by the application of a magnetic field. } 
\TextBlue{In fact, it has been reported that \Tk{spin}} 
in the \TRevD{quasi-one-dimensional} spin system Sr$_2$V$_3$O$_9$ 
\TextBlue{with the intrachain interaction of 82~K is suppressed by the application of a magnetic field of 14~T.}\cite{Uesaka_2010,Kawamata_2014} 
\TextBlue{As for} the behavior of \TextBlue{the} 
\TextRed{thermal conductivity in magnetic fields \TextBlue{at low temperatures} below $T_{\rm N}$\TextBlue{, it} is} 
slightly \TextBlue{complicated}. 


\begin{figure}[b]
\begin{center}
\includegraphics[width=0.8\linewidth]{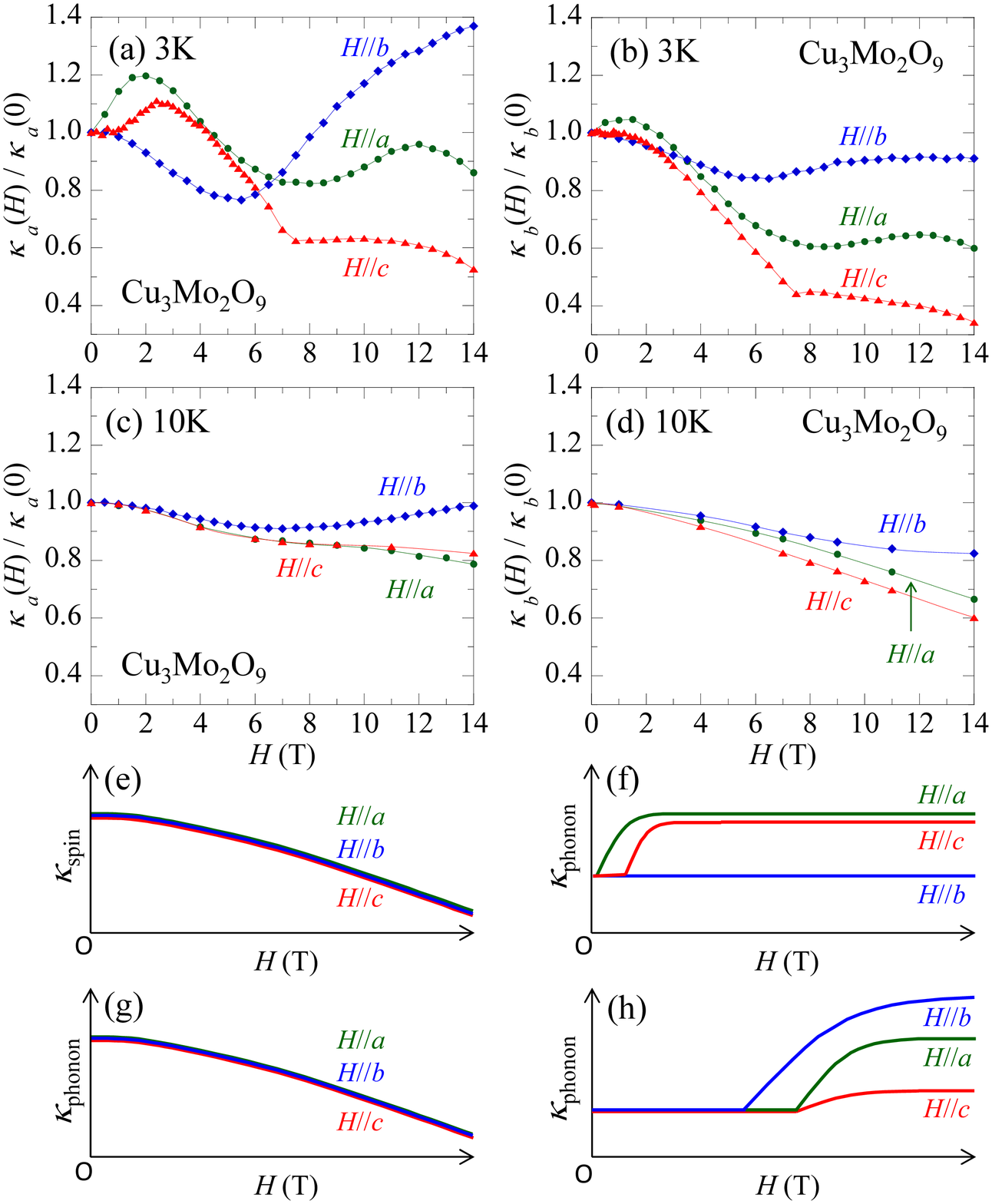}
\vspace{5mm}
\caption{
(Color online) 
\TRevB{(a)--(d)} Magnetic-field dependence of the thermal conductivity along the $a$- and $b$-axes normalized by the value in zero field, $\kappa_{a}(H)/\kappa_{a}(0)$ and $\kappa_{b}(H)/\kappa_{b}(0)$, respectively, for Cu$_3$Mo$_2$O$_9$ single crystals in magnetic fields parallel to the $a$-, $b$-, and $c$-axes at 3 and 10 K. 
\TRevB{(e)--(h) Schematic diagrams of the magnetic-field dependence of the thermal conductivity due to magnetic excitations, \Tk{spin}, and phonons, \Tk{phonon}. 
(e) \Tk{spin} suppressed by the reduction in the spin gap. 
(f) \Tk{phonon} enhanced by the appearance of the long-range order of the canted components in the weak ferromagnetic state. 
(g) \Tk{phonon} suppressed by the reduction in the spin gap. 
(h) \Tk{phonon} enhanced in high magnetic fields. 
}}
\label{fig04}
\end{center}
\end{figure}


Figures \ref{fig04}(a)--\ref{fig04}(d) show \TextBlue{the} \TextRed{magnetic-field dependences of} $\kappa_{a}$($H$)\TextRed{/$\kappa_{a}$(0)}, and $\kappa_{b}$($H$)\TextRed{/$\kappa_{b}$(0)} 
\TextBlue{of Cu$_3$Mo$_2$O$_9$, normalized by the value in zero field, in \THax{a}, \THax{b} and \THax{c}} 
at 3 and 10 K. 
\TextRed{First, we \TextBlue{compare} $\kappa_{a}$($H$){/$\kappa_{a}$(0)} and $\kappa_{b}$($H$){/$\kappa_{b}$(0)}. } 
\TextRed{It is found that both $\kappa_{a}$($H$){/$\kappa_{a}$(0)} and $\kappa_{b}$($H$){/$\kappa_{b}$(0)} show \TextBlue{a complicated but similar behavior in general terms, but} 
$\kappa_{b}$($H$)/$\kappa_{b}$(0) tends to decrease with increasing field \TextBlue{more} than $\kappa_{a}$($H$)/$\kappa_{a}$(0). }
\TextRed{Here, $\kappa_{b}$ parallel to \TextBlue{the} spin chains is described as the sum of \Tk{phonon} and \Tk{spin}, while $\kappa_{a}$ perpendicular to \TextBlue{the} spin chains is \TextBlue{given by} only \Tk{phonon}. }
Therefore, it is inferred that the complicated field-dependence of the thermal conductivity is due to \Tk{phonon}, 
while \Tk{spin} monotonically decreases with increasing field, \TRevB{as shown in Fig. \ref{fig04}(e)}. 

\TextRed{Next, we discuss \TextBlue{the} magnetic-field dependence of $\kappa_{a}$($H$){/$\kappa_{a}$(0)} in order to investigate \TextBlue{the} magnetic and dielectric states, because} 
the behavior of $\kappa_{a}(H)\TextBlue{/\kappa_{a}(0)}$ \TextBlue{originating from} \TextRed{only \Tk{phonon}} 
is expected to reflect \TextRed{these} states \TextRed{through} \TextBlue{the scattering of phonons more simply than that of $\kappa_{b}(H)/\kappa_{b}(0)$}. 
\TextRed{It is found that \TextBlue{the field dependence of} $\kappa_{a}(H)/\kappa_{a}(0)$} 
at 3 K \TextBlue{is very} different \TextBlue{depending on} the \TextBlue{applied-field-}direction, as shown in Fig.~\ref{fig04}(a). 
\TextRed{In low magnetic fields, $\kappa_{a}(H)/\kappa_{a}(0)$ increase\TextBlue{s} \TextBlue{up to} $\sim \TextBlue{2}$~T \TextBlue{with increasing fields} of \THax{a} and \THax{c}, while \TextBlue{it} decrease\TextBlue{s up to $\sim 5$~T with increasing field of \THax{b}}.} 
\TextRed{Since the long-range order of 
\TextBlue{canted components of the magnetic moments in} WF appear\TextBlue{s} above $\THax{a}\sim 0.1$~T and $\THax{c}\sim 0.8$~T but it does not in \THax{b} \cite{Kuroe_2011_JPSJ}, 
the increase \TextBlue{in} $\kappa_{a}(H)/\kappa_{a}(0)$ \TextBlue{with increasing fields of} \THax{a} and \THax{c} is \TextBlue{explained as being} caused by the appearance of the long-range order of \TextBlue{the canted components in WF leading to the suppression of the phonon\TextBlue{-spin} scattering.}} 
 \TRevB{Therefore, there is an enhanced component of \Tk{phonon} in both \THax{a} and \THax{c}, as shown in Fig. \ref{fig04}(f)}. 
\TextRed{The decrease \TextBlue{in} $\kappa_{a}(H)/\kappa_{a}(0)$ in \THax{b} is \TextBlue{explained as being} caused by the \TextBlue{increase} in the phonon\TextBlue{-spin} scattering \TextBlue{rate} due to the \TextBlue{reduction} in \TextBlue{the} spin gap by \TextBlue{the} application of a magnetic field.} 
\TextRed{Furthermore, \TextBlue{it is found that} 
$\kappa_{a}(H)/\kappa_{a}(0)$ start\TextBlue{s} to decrease \TextBlue{above} $\sim 2$~T with increasing fields \TextBlue{of \THax{a} and \THax{c}},
\TextBlue{which is interpreted as being caused by both the} 
saturation of the enhancement of $\kappa_{a}(H)/\kappa_{a}(0)$ by \TextBlue{the} appearance of \TextBlue{the} long-range order of \TextBlue{the canted components in} WF, \TRevB{as shown in Fig.~\ref{fig04}(f)}, 
and \TextBlue{the} decrease \TextBlue{in} $\kappa_{a}(H)/\kappa_{a}(0)$ \TextBlue{due to} the \TextBlue{reduction} in \TextBlue{the} spin gap}, 
\TRevB{as shown in Fig.~\ref{fig04}(g)}.


\TextRed{In high magnetic fields, $\kappa_{a}(H)/\kappa_{a}(0)$ \TextBlue{tends to} increase \TextBlue{above $\sim 7$~T with increasing fields of \THax{a}, \THax{b}, and \THax{c}}, as shown in Fig. \ref{fig04}(a).} 
\TextRed{In particular, \TextBlue{it is remarkable that} 
there is a kink in $\kappa_{a}$($H$)/$\kappa_{a}$(0) at \THax{c} $\sim$7.5 T, where the phase transition occurs, that is, the direction of the spontaneous electric polarization changes from the $c$-axis to the $a$-axis with increasing field, as shown in Fig. 2\cite{Kuroe_2011_JPSJ}. }
\TextBlue{A similar kink is also observed in $\kappa_{b}(H)/\kappa_{b}(0)$ at $\THax{c} \sim 7.5$~T.} 
\TextBlue{In \THax{a} and \THax{b}, on the other hand, no anomaly suggesting any phase transitions has been observed at around 7~T in the specific heat \cite{Hamasaki_2010} and magnetization \cite{Hamasaki_2009} measurements.} 
\TextBlue{However, since} the differential magnetization \TextBlue{has shown} a kink at \THax{b} = 6~T at 2~K, 
the enhancement of \TextRed{$\kappa_{a}(H)/\kappa_{a}(0)$} \TextBlue{above $\sim 7$~T} may be caused by \TextBlue{an unknown} field-induced order \TextRed{and/or \TextBlue{a} change in \TextBlue{the} magnetic state}. 
\TRevB{Accordingly, there is an enhanced component of \Tk{phonon} in \THax{a}, \THax{b}, and \THax{c}, as shown in Fig. \ref{fig04}(h)}. 
\TextRed{The \TextBlue{enhancement of $\kappa_{a}(H)/\kappa_{a}(0)$ above} $\sim 7$~T is \TextBlue{also} observed at 10~K above $T_{\rm N}$, as shown in Fig. \ref{fig04}(c).} 


\begin{figure}[t]
\begin{center}
\includegraphics[width=0.5\linewidth]{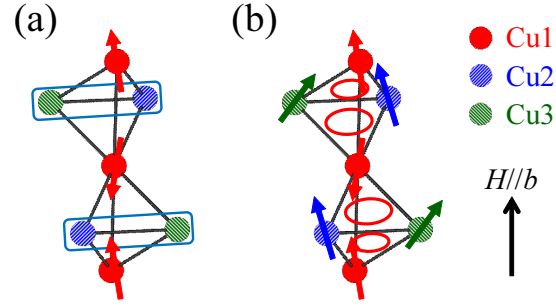}
\caption{
(Color online) 
Schematic diagram of \TextBlue{spin} chiralities and spontaneous current\TextBlue{s} caused by \TextBlue{the break} of \TextBlue{dimers of} Cu2 and Cu3 in Cu$_3$Mo$_2$O$_9$. 
(a) In the case \TextBlue{that} spins of Cu2 and Cu3 form \TextBlue{a spin-}singlet dimer, there is no chirality in \TextBlue{the} spin chain. 
Rounded rectangles  and arrows indicate \TextBlue{spin}-singlet dimer\TextBlue{s} and spins \TextRed{on \TextBlue{the Cu1} site}, respectively. 
(b) In \TextBlue{a} magnetic field along the $b$-axis, \TextRed{spontaneous currents \TextBlue{run} along triangles \TextBlue{composed of} three Cu \TextBlue{spins, owing to the break of spin-singlet dimers.}} 
Open ovals and arrows indicate spontaneous currents and spins, respectively. }
\label{fig05}
\end{center}
\end{figure}

Here, \TextBlue{it is noted that} 
the enhancement of \TextBlue{$\kappa_{a}(H)/\kappa_{a}(0)$} above $\sim 7$~T means the increase of \Tl{phonon}, 
because \TextBlue{both} the specific heat and velocity of phonons are usually almost independent \TextBlue{of} magnetic field. 
In other word\TextBlue{s}, \TextBlue{it means that} the scattering \TextBlue{rate of phonons} decreases with increasing field, 
corresponding to the decrease in magnetic excitations and/or \TextBlue{the} development of \TextBlue{a} magnetic order. 
According to the calculation of the magnetic dispersion in magnetic fields by Matsumoto \textit{et al}.\cite{Matsumoto_2012_JPSJ}, \TextBlue{no anomaly} such as \TextBlue{any} change in the ground state \TextBlue{has been suggested} at $\sim 7$~T. 

\TextBlue{Here,} 
in order to \TextBlue{explain} the enhancement of $\kappa_{\rm phonon}$ \TextBlue{above $\sim 7$~T}, we introduce the theory \TextBlue{proposed by Bulaevskii and Batista \cite{Bulaevskii_2008} and Khomskii \cite{Khomskii_2010}} concerning spontaneous currents \TextRed{and charge redistribution} 
in \TextBlue{a Mott insulator regarded as a} geometrically frustrated spin system. 
Since the ferroelectricity in Cu$_3$Mo$_2$O$_9$ has been \TextBlue{understood} on the basis of the charge redistribution \cite{Kuroe_2011_JPSJ}, the spontaneous currents may be useful to \TextBlue{explain} the enhancement of $\kappa_{\rm phonon}$. 
In \TextBlue{a} geometrically frustrated Mott insulator, \TextBlue{the} exchange interaction \TextBlue{between three spins forming a triangle} cause\TextBlue{s a} spontaneous current running along the triangle. 
This spontaneous current only appear\TextBlue{s in a} non-coplanar spin-state 
\TextRed{and is proportional to }
\TextBlue{the} scalar spin-chirality 
\TextRed{given by ${\boldsymbol S}_1 \cdot ({\boldsymbol S}_2 \times {\boldsymbol S}_3)$, where ${\boldsymbol S}_i$ $\TextBlue{(i = 1, 2, 3)}$ is a spin angular momentum on \TextBlue{the} site $i$}. 
In the case \TextBlue{that} the spins of Cu2 and Cu3 form \TextBlue{a spin-}singlet dimer, \TextBlue{distorted} tetrahedral spin-chains can be regarded as simple spin-chains composed \TextBlue{of only} Cu1 spins and there is no chirality in \TextBlue{the} spin chain\TextBlue{,} as shown in Fig. \ref{fig05}(a). 
\TextRed{In the case that \TextBlue{spin-}singlet dimers are \TextBlue{broken} by the application of a magnetic field,} \TextBlue{on the other hand, finite values of spin chirality} appear \TextBlue{in the triangles}, 
because \TextRed{\TextBlue{spins revive on the} Cu2 and Cu3 sites, as shown} in Fig. \ref{fig05}(b). 
\TextBlue{Therefore, it is possible that} the enhancement of $\kappa_{\rm phonon}$ \TextBlue{above $\sim 7$~T is} caused by the ordering \TextBlue{of spin chiralities}, because the ordering \TextBlue{is able to} be \TextBlue{brought about} by \TextBlue{the} magnetic interaction even in the absence of \TextBlue{any} magnetic\TextBlue{ally} ordered state. 

\begin{figure}[t]
\begin{center}
\includegraphics[width=1.0\linewidth]{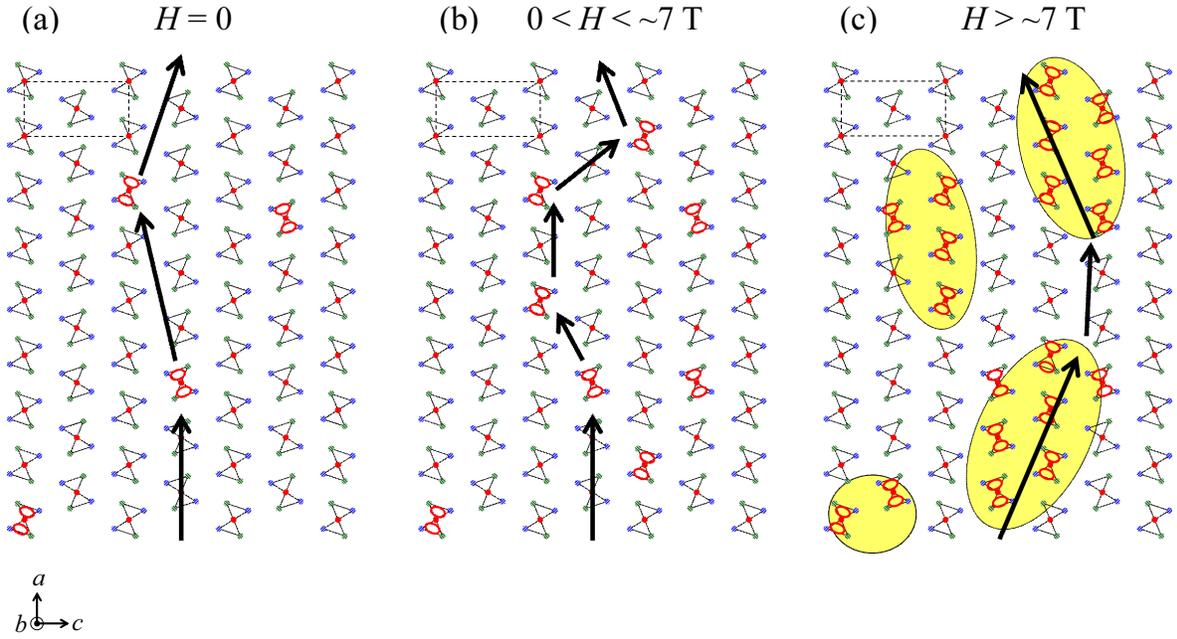}
\caption{
(Color online) 
Schematic diagram of the development of the \TextBlue{spin-}chirality order \TextBlue{in Cu$_3$Mo$_2$O$_9$}. 
Open ovals and the length of arrows indicate excitations of \TextBlue{spin-singlet} dimers and \TextBlue{the} mean free path of phonons, $l_{\rm phonon}$, respectively. 
(a) \TextRed{In zero field, }
the magnitude of $l_{\rm phonon}$ is limited by magnetic excitations \TextRed{generated by thermal fluctuations}. 
(b) In \TextRed{low} magnetic fields \TextRed{below $\sim 7$~T}, 
the number of magnetic excitations increases \TextBlue{with increasing field} because of the \TextBlue{reduction} in the spin gap\TextBlue{, so that} 
$l_{\rm phonon}$ is \TextBlue{shortened} because of the \TextBlue{increase} in the phonon scattering \TextBlue{rate}. 
(c) The \TextBlue{spin-}chirality order \TextBlue{is developed} above \TextRed{$\sim 7$~T}\TextBlue{, so that} $l_{\rm phonon}$ extends because of the \TextBlue{decrease} in the phonon scattering \TextBlue{rate}. 
Filled ovals indicate \TextOliveGreen{areas of the spin-chirality order}.}
\label{fig06}
\end{center}
\end{figure}

\TextBlue{Finally, the magnetic-field dependence of $\kappa_{a}(H)/\kappa_{a}(0)$ at low temperatures below $\sim 40$~K is summarized as follows, on the basis of the scenario adopting the spin-chirality ordering.} 
In zero field, a few excitations 
\TextBlue{of spin-singlet dimers in the spin-gap state} 
due to thermal fluctuation\TextBlue{s} \TextRed{scatter phonons, as shown in} Fig. \ref{fig06}(a). 
\TextRed{Since} the number of magnetic excitations increases \TextRed{with increasing field below $\sim 7$~T} by the \TextBlue{reduction} in the spin gap, 
$l_{\rm phonon}$ \TextBlue{is shortened} because of the \TextBlue{increase} in the phonon scattering \TextBlue{rate,} \TextRed{as shown in} Fig. \ref{fig06}(b). 
\TextRed{In} \TextBlue{high} magnetic fields above \TextRed{$\sim 7$~T}, 
\TextBlue{the} order of magnetic excitations\TextBlue{, namely, the order of spin chiralities, is developed,} \TextRed{as shown in } Fig. \ref{fig06}(c), 
\TextBlue{so that} 
\TextRed{$\kappa_{\rm phonon}$ \TextBlue{increases owing to the} decrease in \TextBlue{the} phonon scattering \TextBlue{rate}.} 
The reason why \TextBlue{the} \TextRed{enhancement of} 
\TextBlue{$\kappa_{a}(H)/\kappa_{a}(0)$} \TextRed{is \TextBlue{different depending on the applied-field-}direction is as follows.} 
Spontaneous currents \TextRed{along \TextBlue{the} triangles \TextBlue{composed of three spins} induce } orbital moments, which \TextBlue{are} coupled \TextBlue{with} the magnetic field. 
Therefore, \TextRed{the magnitude of} \TextBlue{the} scalar spin-chirality \TextRed{might be related to} the magnetic field \TextBlue{penetrating} the triangles. 
\TextBlue{Accordingly,} \TextRed{since} the area\TextBlue{s} of the triangles viewed from the $b$-axis \TextBlue{are} \TextRed{homogeneous}, the chirality order may \TextRed{be} homogeneous in \TextRed{\THax{b}, leading \TextBlue{to} the large enhancement of $\kappa_{\rm phonon}$}. 
\TextRed{On the other hand, since} the area\TextBlue{s} of the triangles viewed from the $a$- and $c$-axes \TextBlue{are} \TextRed{inhomogeneous}, the chirality order may \TextRed{be} inhomogeneous in \TextBlue{\THax{a} and \THax{c},} \TextRed{leading \TextBlue{to} the small enhancement of $\kappa_{\rm phonon}$}. 
\TextBlue{To confirm} this \TextBlue{scenario adopting the spin-chirality} order, further experimental and theoretical investigations \TextBlue{are necessary}.

\section{\TextBlue{Summary}}
In order to investigate the magnetic state 
\TextBlue{and the existence of \Tk{spin}}, 
we have measured \TextRed{\Tkax{a}, \Tkax{b}, and \Tkax{c}} of \TextBlue{Cu$_3$Mo$_2$O$_9$} single crystals in magnetic fields \TextRed{up to 14~T}. 
\TextRed{In zero field, it has been found that} 
\TextBlue{\Tkax{a}, \Tkax{b}, and \Tkax{c} are suppressed at high temperatures probably by magnetic fluctuations due to the spin frustration, 
while they are enhanced just below \TT{N} as in the case of several antiferromagnets.} 
\TextRed{By the application of a magnetic field, }
\TextRed{\Tkax{a}, \Tkax{b}, and \Tkax{c} \TextBlue{have been found to be} suppressed at low temperatures below $\sim 40$~K }
and this \TextBlue{has been explained as being due to} 
\TextRed{the reduction in \TextBlue{the} spin gap \TextBlue{originating from} the \TextBlue{spin-}singlet dimer\TextBlue{s of} Cu2 and Cu3. }
\TextBlue{Since it has been found that the magnitude of \Tkax{b} parallel to the spin chains is larger than those of \Tkax{a} and \Tkax{c} and that the decrease in \Tkax{b} by the application of a magnetic field is more marked than that in \Tkax{a}, it is concluded that there exists a contribution of \Tk{spin} to \Tkax{b}.}
\TextRed{Furthermore, 
\TextBlue{it has been found that} 
the magnetic-field dependences of \Tkax{a} and \Tkax{b} at 3 and 10~K are \TextBlue{complicated} and different} \TextBlue{depending on the applied-field-direction}. 
\TextBlue{In low magnetic fields below $\sim 7$~T, 
both \Tkax{a} and \Tkax{b} have been found to decrease with increasing field due to the reduction in the spin gap. 
Moreover, \Tkax{a} at 3~K has been found to markedly change in \THax{a} and \THax{c} in correspondence to the appearance of the long-range order of the canted components in WF. 
In high magnetic fields above $\sim 7$~T, on the other hand, both \Tkax{a} and \Tkax{b} at 3~K have been found to tend to increase with increasing field. 
In \THax{c}, a kink has been observed at $\sim 7.5$~T in both \Tkax{a} and \Tkax{b}, owing to the field-induced phase transition. 
In \THax{b}, it has been found that the increase in \Tkax{a} above $\sim 7$~T is most marked and is observed even at 10~K above \TT{N} in spite of the absence of any phase transition, suggesting the existence of a novel field-induced spin state. }
\TextBlue{A} possible \TextBlue{state} is the \TextBlue{ordered one of the spin} chirality in \TextBlue{a} frustrated Mott insulator.

\section*{Acknowledgment}


The thermal conductivity measurements were performed at the High Field Laboratory for Superconducting Materials, Institute for Materials Research, Tohoku University. 
\TextOliveGreen{Figure~1 was drawn using VESTA \cite{Momma_2011}. }
This work was supported by a Grant-in-Aid for Scientific Research from the Ministry of Education, Culture, Sports, Science and Technology, Japan.


\end{document}